\documentclass[a4paper,11pt]{article}
\usepackage[utf8]{inputenc}
\usepackage[T1]{fontenc}
\usepackage[english]{babel}
\usepackage[hmargin={32mm,32mm},vmargin={32mm,35mm}]{geometry}
\usepackage{amsmath}
\usepackage{amsfonts}
\usepackage{amssymb}
\usepackage[mathscr]{euscript}
\usepackage{enumerate}
\usepackage{graphicx}
\usepackage{bbm}
\usepackage{color}
\usepackage{xspace}
\usepackage{url}
\usepackage[hidelinks]{hyperref}
\usepackage{caption}
\usepackage{enumitem}
\usepackage{bookmark}
\usepackage{csquotes}
\usepackage[
	bibstyle=phys,
	biblabel=brackets,
	citestyle=numeric-comp,
	sorting=none,
	doi=false,
	eprint=true,
	pageranges=false,
	maxbibnames=10,
	backend=biber
]{biblatex}

\DeclareFieldFormat[article,inproceedings,inbook]{title}{\mkbibitalic{#1\isdot}}
\AtEveryBibitem{
	\ifentrytype{book}{
		\clearfield{series}
		\clearfield{volume}
	}{}
	\ifentrytype{inbook}{
		\clearfield{pages}
	}{}
}

\newcommand{\bra}[1]{\langle #1 |}
\newcommand{\ket}[1]{|#1\rangle}

\newcommand{\updown}[2]{^{#1}_{\phantom{#1}#2}}

\newcommand{\Tr}{\operatorname{Tr}}
\newcommand{\D}{{\mathscr{D}}}
\newcommand{\V}{{\cal V}}

\newcommand{\threedots}{{\lower.25em\hbox{\vdots}}}

\numberwithin{equation}{section}

\begin{document}

\begin{center}

\Large
\textbf{Quantum-reduced loop gravity:\\New perspectives on the kinematics and dynamics}

\vspace{16pt}

\large
Ilkka Mäkinen

\normalsize

\vspace{12pt}

National Centre for Nuclear Research \\
Pasteura 7, 02-093 Warsaw, Poland

\vspace{8pt}
 
ilkka.makinen@ncbj.gov.pl

\end{center}

\renewcommand{\abstractname}{\vspace{-\baselineskip}}

\begin{abstract}
	\noindent We present a systematic approach to the kinematics of quantum-reduced loop gravity, a model originally proposed by Alesci and Cianfrani as an attempt to probe the physical implications of loop quantum gravity. We implement the quantum gauge-fixing procedure underlying quantum-reduced loop gravity by introducing a master constraint operator on the kinematical Hilbert space of loop quantum gravity, representing a set of gauge conditions which classically constrain the densitized triad to be diagonal. The standard Hilbert space of quantum-reduced loop gravity can be recovered as a space of solutions of the master constraint operator, while on the other hand the master constraint approach provides a useful starting point for considering possible generalizations of the standard construction. We also examine the quantum dynamics of states consisting of a single six-valent node in the quantum-reduced framework. We find that the Hamiltonian which governs the dynamics of such states bears a close formal resemblance to the Hamiltonian constraint of Bianchi I models in loop quantum cosmology.
\end{abstract}

{\let\thefootnote\relax\footnotetext{\hspace{-20.5pt} This article is a contribution to the proceedings of the 17th Marcel Grossmann Meeting (Pescara, Italy, 7-12 July 2024).}}

\section{Introduction}

Quantum-reduced loop gravity is a simplified model of loop quantum gravity, introduced by Alesci and Cianfrani in 2013 \cite{Alesci:2012md, Alesci:2013xd}. While the original motivation for Alesci and Cianfrani's work was was primarily to shed light on the relation between loop quantum gravity and loop quantum cosmology \cite{Alesci:2014rra}, their model has since found a wider range of physical applications related to cosmology \cite{Alesci:2017kzc, Alesci:2018qtm, Olmedo:2018ohq, Alesci:2019sni} as well as black holes \cite{Alesci:2019pbs, Alesci:2020zfi, Gan:2022mle}.

At the heart of quantum-reduced loop gravity is the notion of quantum gauge fixing: A set of gauge conditions -- in this case fixing the densitized triad to be diagonal -- is used at the level of the quantum theory to select a sector of the full Hilbert space of loop quantum gravity, which then serves as the Hilbert space of the quantum-reduced model. In the present work \cite{Makinen:2023shj} we present an alternative point of view on the kinematical foundations of quantum-reduced loop gravity by proposing the master constraint method \cite{Thiemann:2003zv, Thiemann:2005zg} as a natural and perhaps conceptually cleaner approach to implement the gauge fixing procedure underlying the quantum-reduced model. We will also see a concrete example of how the master constraint approach provides a fruitful standpoint for thinking about generalizations of the standard framework of quantum-reduced loop gravity.

As a complementary line of investigation, we study the quantum dynamics (as opposed e.g.~to effective dynamics based on semiclassical expectation values of the Hamiltonian) of simple states of quantum geometry in the quantum-reduced setting. Considering a model in which the entire universe is represented by a state consisting of a single six-valent vertex, we find that the dynamics of such a state is governed by a Hamiltonian operator whose form is very similar to the Hamiltonian constraint of Bianchi I models in loop quantum cosmology. Once understood in more detail, this result may bring a new perspective to the question of how loop quantum cosmology and full loop quantum gravity are related to each other.

\section{Elements of loop quantum gravity}
\label{sec:lqg}

Our work takes place in the standard kinematical setting of loop quantum gravity \cite{Ashtekar:2004eh, Rovelli:2004tv, Thiemann:2007pyv, Rovelli:2014ssa, Ashtekar:2017yom}. We therefore begin begin by giving a very brief review of the most essential elements of the framework in this section.

The kinematical Hilbert space ${\cal H}_{\rm kin}$ of loop quantum gravity is spanned by the so-called cylindrical functions. Let $\Gamma$ be a graph consisting of $N$ oriented edges $e_1, \dots, e_N$ embedded in the spatial manifold $\Sigma$. Then a function cylindrical with respect to $\Gamma$ is a complex-valued function which depends on $N$ group elements of $SU(2)$, these group elements being associated with the edges of the graph $\Gamma$. A scalar product on the space of cylindrical functions is given by the Ashtekar--Lewandowski scalar product \cite{Ashtekar:1993wf, Ashtekar:1994mh}, which is defined in a natural way using the $SU(2)$ Haar measure.

An orthonormal basis on the space of functions cylindrical with respect to a given graph $\Gamma$ is formed by the functions
\begin{equation}
	\prod_{e\in\Gamma} \D^{(j_e)}_{m_en_e}(h_e),
	\label{}
\end{equation}
where the $SU(2)$ quantum numbers range over all their possible values, and we have introduced the notation
\begin{equation}
	\D^{(j)}_{mn}(h) = \sqrt{2j+1} D^{(j)}_{mn}(h)
	\label{}
\end{equation}
to denote the normalized matrix elements of the $SU(2)$ representation matrices. Although they will not play a very important role in this work, we may also mention the well-known spin network functions \cite{Rovelli:1995ac, Baez:1995md}, which span the $SU(2)$ invariant subspace of ${\cal H}_{\rm kin}$. These are cylindrical functions of the form
\begin{equation}
	\biggl(\prod_{v\in\Gamma} \iota_v\biggr) \cdot \biggl(\prod_{e\in\Gamma} \D^{(j_e)}(h_e)\biggr),
	\label{eq:spinnetwork}
\end{equation}
where each node of the graph carries an intertwiner (invariant tensor) $\iota_v$, whose index structure mirrors the number and orientation of the edges incident on the node, and the dot denotes a complete contraction of the magnetic indices.

Let us then introduce the elementary operators on ${\cal H}_{\rm kin}$. The holonomy operator associated to an edge $e$ (which may or may not be an edge of the graph $\Gamma$) acts on cylindrical functions as a multiplicative operator:
\begin{equation}
	\widehat{D^{(j)}_{mn}(h_e)}\Psi_\Gamma(h_{e_1}, \dots, h_{e_N}) = D^{(j)}_{mn}(h_e)\Psi_\Gamma(h_{e_1}, \dots, h_{e_N}).
	\label{eq:D-def}
\end{equation}
The role of the conjugate momentum operator is essentially fulfilled by the spin operator $\hat J_i^{(v, e)}$, which carries an $su(2)$ algebra index $i$ and is labeled by a point $v$ and an edge $e$. Its action on a representation matrix is defined as
\begin{equation}
	\hat J_i^{(v, e)}\D^{(j)}(h_e) = \begin{cases}
		i\D^{(j)}(h_e)\tau^{(j)}_i & v = s(e) \\
		-i\tau^{(j)}_i\D^{(j)}(h_e) & v = t(e) \\
	\end{cases}
	\label{eq:J-def}
\end{equation}
where $s(e)$ and $t(e)$ refer to the starting and ending points (source and target) of the edge $e$, and $\tau_i = -i\sigma_i/2$ are the anti-Hermitian generators of $SU(2)$. The action \eqref{eq:J-def} is extended to arbitrary cylindrical functions by linearity and the Leibniz rule. The holonomy operator and the spin operator are the fundamental building blocks out of which more elaborate operators, such as those representing geometrical quantities like areas, volumes and others, as well as those governing the dynamics of the quantum states of loop quantum gravity, can be constructed.

\section{Master constraint for diagonal gauge}

We wish to construct a constraint operator on the kinematical Hilbert space, which encodes the gauge conditions enforcing the densitized triad $E^a_i$ to be diagonal, i.e.
\begin{equation}
	E^a_i = 0 \qquad (a\neq i)
	\label{eq:gauge}
\end{equation}
(We assume that spatial indices and internal indices take the values $a = x, y, z$ and $i = 1, 2, 3$ respectively, but the values $1, 2, 3$ are identified, and can be used interchangeably, with $x, y, z$.) To carry out this task, we will employ the master constraint method, which was first introduced to loop quantum gravity by Thiemann in connection with the Hamiltonian constraint \cite{Thiemann:2003zv, Thiemann:2005zg}. We begin by defining the classical constraint function
\begin{equation}
	\mu^2 = (E^x_2)^2 + (E^x_3)^2 + (E^y_1)^2 + (E^y_3)^2 + (E^z_1)^2 + (E^z_2)^2
	\label{eq:mu2}
\end{equation}
such that the condition $\mu^2(x) = 0$ is equivalent to all the gauge conditions \eqref{eq:gauge} being satisfied at the point $x$. We then introduce the integrated master constraint
\begin{equation}
	M = \int d^3x\, \frac{\mu^2}{\sqrt q}
	\label{eq:M}
\end{equation}
where we have divided by the volume element $\sqrt q$ in order to obtain an integrand which has the appropriate density weight of $+1$. Since the integrand is a positive definite function, the constraint equation $M=0$ is satisfied if and only if the gauge conditions \eqref{eq:gauge} hold at every point on the spatial manifold $\Sigma$.

In order to pass from the classical functional \eqref{eq:M} into a quantum operator on the kinematical Hilbert space, we assume that a fixed Cartesian background coordinate system is given, and make use of certain structures defined with respect to this coordinate system \cite{Makinen:2023shj}. In this way we obtain an operator whose action on a cylindrical function takes the form
\begin{equation}
	\hat M = \sum_v \hat M_v
	\label{}
\end{equation}
with
\begin{equation}
	\hat M_v = \widehat{\V_v^{-1}}\sum_a \Bigl(\hat A\bigl(S^a(v)\bigr)^2 - \hat E_a\bigl(S^a(v)\bigr)^2\Bigr).
	\label{eq:M_v}
\end{equation}
Here $S^a(v)$ $(a = x, y, z)$ is a surface which intersects the graph of the cylindrical function at the node $v$ (and not at any other node), and is dual to the $x^a$ -coordinate direction (i.e.~the coordinate $x^a$ is constant on the surface). Moreover, $\hat E_i(S)$ denotes the flux operator, while
\begin{equation}
	\hat A(S) = \sqrt{\displaystyle \hat E^i(S) \hat E_i(S)}
	\label{}
\end{equation}
is the area operator associated to the surface $S$. Finally,
\begin{equation}
	\widehat{\V_v^{-1}} := \lim_{\epsilon\to 0} \frac{\hat V_v}{\hat V_v^2 + \epsilon^2}
	\label{}
\end{equation}
is the so-called Tikhonov regularization of the inverse of the local volume operator $\hat V_v$ at the node $v$. (In terms of its spectral decomposition, this operator can be characterized as follows: Let $\ket\lambda$ be an eigenstate of $\hat V_v$ with eigenvalue $\lambda$. Then $\widehat{\V_v^{-1}}\ket\lambda = \lambda^{-1}\ket\lambda$ or $0$, according to whether the eigenvalue $\lambda$ is non-zero or zero.) This concludes the construction of the master constraint operator representing the classical gauge conditions \eqref{eq:gauge} in the quantum theory.

\section{The Hilbert space of quantum-reduced loop gravity}

The constraint operator for diagonal gauge having been defined, the next step is to search for solutions of the constraint, i.e.~states satisfying the constraint equation
\begin{equation}
	\hat M\ket\Psi = 0.
	\label{eq:Mpsi}
\end{equation}
It is worth noting that the gauge conditions \eqref{eq:gauge} do not commute with each other as operators in the quantum theory, in spite of being classically compatible with each other. On the technical level, this is reflected in the fact that the factor $\sum_a \bigl(\hat A(S^a)^2 - \hat E_a(S^a)^2\bigr)$ in Eq.~\eqref{eq:M_v} is a strictly positive operator. Hence we take the point of view that it would be too restrictive to insist on imposing Eq.~\eqref{eq:Mpsi} as an exact equality. Instead, following the same approach which Alesci and Cianfrani adopted when originally introducing quantum-reduced loop gravity, we look for solutions of the constraint equation among states carrying large spin quantum numbers, and interpret Eq.~\eqref{eq:Mpsi} as an equality which is to be approximately satisfied in the regime of large spins, and which becomes exact only in the formal limit $j\to\infty$.

A particular family of solutions can be obtained by taking as an ansatz a state defined on a cubical graph, whose edges are aligned with the coordinate axes of the previously introduced background coordinate system. In order to write down the resulting solutions, it is useful to introduce the notation $\ket{jm}_i$ to denote the eigenstates of the operators $\hat J^2$ and $\hat J_i$ $(i = x, y, z)$, and
\begin{equation}
	\D^{(j)}_{mn}(h)_i = {}_i\bra{jm}\D^{(j)}(h)\ket{jn}_i
	\label{}
\end{equation}
for the matrix elements of the (normalized) Wigner matrices in the basis $\ket{jm}_i$. Then a space of solutions of Eq.~\eqref{eq:Mpsi} (in the approximate sense discussed above) is given by the states \cite{Makinen:2023shj}
\begin{equation}
	\prod_{e\in\Gamma} \D^{(j_e)}_{\tau_ej_e\; \sigma_ej_e}(h_e)_{i_e}
	\label{eq:basis_reduced}
\end{equation}
where $\Gamma$ is a cubical graph, and the following conditions are assumed to be satisfied:
\begin{itemize}
	\item $j_e \gg 1$ for every edge $e$;
	\item The label $i_e = x, y, z$ is chosen according to the direction of the edge $e$;
	\item The sign factors $\sigma_e$ and $\tau_e$ (belonging respectively to the source and target of the edge $e$) take the values $\pm 1$. The factors associated to a given edge $e$ are independent of each other, but the factors on consecutive edges must be matched in the following sense: If $e_a(v)$ and $e_a'(v)$ aligned in the $x^a$ -coordinate direction and meeting each other at a node $v$, with $e_a(v)$ ending at $v$ and $e_a'(v)$ beginning from $v$, then the condition
		\begin{equation}
			\tau_{e_i(v)} = \sigma_{e_i'(v)}
			\label{eq:signs}
		\end{equation}
	must be satisfied by the sign factors.
\end{itemize}
Apart from the constraint \eqref{eq:signs} on the sign factors, the states \eqref{eq:basis_reduced} are essentially identical with the ``reduced spin network states'' previously considered in the literature of quantum-reduced loop gravity. In terms of the $SU(2)$ quantum numbers, the characteristic properties of these states are the large spins, and the extremal (maximal or minimal) values of the magnetic quantum numbers (with respect to the appropriate basis) on every representation matrix. In comparison with the spin network states \eqref{eq:spinnetwork} of the full theory, perhaps the most distinguishing feature of the reduced spin network states is the absence of any non-trivial intertwiner structure at the nodes.

\section{Operators on the reduced Hilbert space}

The kinematical Hilbert space of quantum-reduced loop gravity is spanned by the basis states \eqref{eq:basis_reduced}. The basic kinematical framework of the model is then completed by the definition of the elementary operators on the reduced Hilbert space. These operators can generally be derived from the corresponding operators of full loop quantum gravity by the following procedure \cite{Makinen:2020rda}: Take an operator $\hat{\cal O}$ of the full theory, apply it to the reduced basis states \eqref{eq:basis_reduced} and identify the term of leading order in the spins. For most (although not all) operators of interest, the leading term is again an element of the reduced Hilbert space, i.e.~the action of the operator has the form
\begin{equation}
	\hat{\cal O}\ket{\Psi_0} = f(j)\ket{\Psi} + g(j)\ket{\Phi}
	\label{eq:Opsi}
\end{equation}
where $\ket{\Psi_0}$ and $\ket\Psi$ are normalized states in the reduced Hilbert space, $\ket\Phi$ is a normalized state generally lying outside of the reduced Hilbert space, and for large spins,
\begin{equation}
	f(j) \gg g(j).
	\label{}
\end{equation}
(The notation here is somewhat schematic; in general $j$ in the above expressions should be understood as a collective label for all the spins entering the matrix elements of the operator.) In the case represented by Eq.~\eqref{eq:Opsi}, one can define a ``reduced'' operator $\hat{\cal O}_R$ by dropping the terms of lower order in $j$. Hence the action of the reduced operator on the reduced basis states is given by
\begin{equation}
	\hat{\cal O}^R\ket{\Psi_0} := f(j)\ket{\Psi}.
	\label{}
\end{equation}
From the point of view of the full theory, the reduced operator serves as a good approximation of the full operator $\hat{\cal O}$ on the state $\ket{\Psi_0}$:
\begin{equation}
	\frac{\bigl|\bigl|\hat{\cal O}\ket{\Psi_0} - \hat{\cal O}^R\ket{\Psi_0}\bigr|\bigr|}{\bigl|\bigl|\hat{\cal O}\ket{\Psi_0}\bigr|\bigr|} \ll 1.
	\label{}
\end{equation}
On the other hand, since the reduced Hilbert space is preserved by the action of the operator $\hat{\cal O}^R$, this operator is a legitimate and well-defined operator from the point of view of the quantum-reduced model. Moreover, the structure of a given reduced operator is typically very simple in comparison with the corresponding operator of the full theory, which represents an important practical advantage of quantum-reduced loop gravity in the context of concrete physical applications.

\section{Extended reduced Hilbert space}

The elementary operators of quantum-reduced loop gravity can be obtained by applying the procedure outlined in the previous section to the elementary operators of the full theory introduced in Sec.~\ref{sec:lqg}. For the reduced holonomy operator $\widehat{D^{(j)}_{mn}(h_e)}{}^R$ one finds that only the diagonal $(m=n)$ and ``anti-diagonal'' $(m = -n)$ components of the operator have a non-vanishing action on the basis states \eqref{eq:basis_reduced} \cite{Makinen:2020rda, Makinen:2023shj}. This may seem somewhat unexpected in light of the fact that the gauge conditions \eqref{eq:gauge} classically do not place any constraints on any components of the Ashtekar connection $A_a^i$. One should however keep in mind that the space spanned by the states \eqref{eq:basis_reduced} is certainly not the most general space of solutions of the gauge-fixing master constraint. Hence the preceding observation about the reduced holonomy operator does not indicate any fundamental problem or inconsistency in the quantum-reduced framework. On the other hand, it does strongly suggest that the standard Hilbert space of quantum-reduced loop gravity is not large enough to capture all relevant configurations where the gauge conditions \eqref{eq:gauge} are satisfied, and that a suitable extension or generalization of this space is needed in order to correctly describe geometries in which the Ashtekar connection is not diagonal.

Let us briefly sketch one possible way of constructing such a generalization (the construction is described in more detail in \cite{Makinen:2023shj}). We begin with the standard basis states \eqref{eq:basis_reduced} and divide each edge $e$ into two non-overlapping segments $e_1$ and $e_2$ (which together span the entire edge $e$). In the basis function \eqref{eq:basis_reduced} we then replace the representation matrix associated to each edge with a product of factors associated to the two segments as follows:
\begin{equation}
	\D^{(j_e)}_{\tau_e j_e\;\sigma_e j_e}(h_e)_{i_e} \; \to \; \D^{(j_e)}_{\tau_e j_e\;m_e}(h_{e_2})_{i_e}\D^{(j_e')}_{m_e'\;\sigma_e j_e'}(h_{e_1})_{i_e}.
	\label{}
\end{equation}
Here $m_e$ and $m_e'$ are additional quantum numbers labeling the generalized basis state; they reside at a degenerate two-valent node in the interior of the original edge $e$, where the segments $e_1$ and $e_2$ meet. In contrast, the quantum numbers located at the proper six-valent nodes of the cubical graph are left unaltered. Since the constraint operator \eqref{eq:M_v} does not act on two-valent nodes due to the presence of the inverse volume operator, this guarantees that the generalized basis states
\begin{equation}
	\prod_{e\in\Gamma} \D^{(j_e)}_{\tau_ej_e\;m_e}(h_{e_2})_{i_e} \, \D^{(j_e')}_{m_e'\;\sigma_ej_e'}(h_{e_1})_{i_e}
	\label{eq:basis_generalized}
\end{equation}
are still valid solutions of the gauge-fixing master constraint, in the same approximate sense as the standard basis states \eqref{eq:basis_reduced}, as long as the conditions listed in the text following Eq.~\eqref{eq:basis_reduced} are satisfied. Moreover, direct calculation shows that the space spanned by these states is preserved by the action of the reduced holonomy operator, and that all components of the operator generally have a non-vanishing action on this space.

\section{Quantum dynamics}

The simplest possible example of dynamics of quantum states in quantum-reduced loop gravity is obtained by considering states which consist of a single six-valent node \cite{Makinen:2024}. Moreover, we assume periodic boundary conditions, or that the spatial manifold is topologically a torus, so the graph on which the state is based essentially comprises only three edges, carrying three independent spin quantum numbers.

Focusing for now on the Euclidean part of the Hamiltonian only, the dynamics is governed by the quantum operator corresponding to the classical functional
\begin{equation}
	C_E(N) = \int d^3x\, N\frac{\epsilon\updown{ij}{k}E^a_iE^b_jF_{ab}^k}{\sqrt{|\det E|}}.
	\label{}
\end{equation}
(The contribution of the Lorentzian part to the dynamics, based on the curvature operator proposed in \cite{Lewandowski:2021iun, Lewandowski:2022xox}, is currently under investigation.) The operator corresponding to the Hamiltonian constraint can be interpreted either as a constraint operator in the vacuum theory \cite{Thiemann:1996aw, Lewandowski:2014hza} or (with $N=1$) as a physical Hamiltonian in a deparametrized context, where an irrotational dust field is employed as a relational time variable \cite{Husain:2011tk, Swiezewski:2013jza, Assanioussi:2017tql}. As a concrete example, we consider the operator \cite{Alesci:2015wla, Assanioussi:2015gka}
\begin{equation}
	\hat C_E(N)\ket{\Psi_\Gamma} = \sum_{v\in\Gamma} N(v)\hat C_E^v\ket{\Psi_\Gamma}
	\label{}
\end{equation}
where
\begin{equation}
	\hat C_E^v = -2\sum_{e \nparallel e' \; \text{at} \; v} \epsilon\updown{ij}{k} \Tr\bigl(\tau_k\widehat{h_{\alpha_{ee'}}}\bigr)\hat J_i^{(v, e)}\hat J_j^{(v, e')} \widehat{\V_v^{-1}}.
	\label{eq:C_v}
\end{equation}
Here the sum runs over all non-tangential pairs of edges at the node $v$, and in the setting of quantum-reduced loop gravity, we take the loop $\alpha_{ee'}$ to be the minimal graph-preserving loop spanned by the edges $e$ and $e'$ together with two additional edges of the cubical graph. We have also omitted an undetermined multiplicative factor, which is left over from quantization as a regularization ambiguity, and, in the example considered here, enters Eq.~\eqref{eq:C_v} simply as an overall multiplicative factor. Here it would seem natural to fix this factor by the requirement that the operator obtained below in Eq.~\eqref{eq:C_R} has the correct semiclassical limit (in the sense of expectation values in coherent states).

Computing the action of the operator \eqref{eq:C_v} on the state
\begin{equation}
	\ket{j_xj_yj_z} = D^{(j_x)}_{j_xj_x}(h_{e_x}) D^{(j_y)}_{j_yj_y}(h_{e_y}) D^{(j_z)}_{j_zj_z}(h_{e_z}),
	\label{}
\end{equation}
we find that the resulting reduced operator takes the form
\begin{align}
	&\bigl(\hat C_E^v\bigr){}^R\ket{j_xj_yj_z} \notag \\
	&= -\frac{8}{\sqrt{j_xj_yj_z}}\Bigl(j_xj_y\hat s(e_x)\hat s(e_y) + j_xj_z\hat s(e_x)\hat s(e_z) + j_yj_z\hat s(e_y)\hat s(e_z)\Bigr)\ket{j_xj_yj_z}
	\label{eq:C_R}
\end{align}
where the operator $\hat s(e)$ is defined by its action on the basis states as
\begin{equation}
	\hat s(e)\ket{j_e} := \frac{1}{2i}\bigl(\ket{j_e+1} - \ket{j_e-1}\bigr).
	\label{}
\end{equation}
It is interesting to compare the operator \eqref{eq:C_R} with the Hamiltonian constraint of Bianchi I models in loop quantum cosmology \cite{Chiou:2006qq, Szulc:2008ar, Martin-Benito:2008dfr, Ashtekar:2009vc}. The classical phase space of Bianchi I is parametrized by canonical variables $c_i$ and $p_i$ $(i = 1, 2, 3)$, which roughly correspond to the Ashtekar connection and the densitized triad. The Euclidean Hamiltonian constraint classically takes the form
\begin{equation}
	C_E^{\rm BI} = -\frac{2}{\sqrt{p_1p_2p_3}}\bigl(p_1p_2c_1c_2 + p_1p_3c_1c_3 + p_2p_3c_2c_3\bigr).
	\label{eq:C_BI}
\end{equation}
In the quantum theory, a basis of the Hilbert space is provided by the states $\ket{p_1, p_2, p_3}$, which are eigenstates of the triad operators $\hat p_i$. The factors of $c_i$ in the Hamiltonian constraint are quantized by first making the replacement $c_i \to \sin(\mu_i c_i)/\mu_i$, where $\mu_i$ is essentially a regularization parameter, and then promoting $\sin\mu_i c_i$ into an operator as
\begin{equation}
	\widehat{\sin \mu_i c_i} = \frac{1}{2i}\Bigl(\widehat{e^{i\mu_i c_i}} - \widehat{e^{-i\mu_i c_i}}\Bigr)
	\label{}
\end{equation}
where the action of the ``holonomy operator'' on the basis states is given by
\begin{equation}
	\widehat{e^{i\mu c}}\ket p = \ket{p-\mu}.
	\label{}
\end{equation}
Hence we see that the operator resulting from such a quantization is virtually identical with the operator \eqref{eq:C_R}, provided that the constant value $\mu_i = 1$ is used for the polymerization parameters, and that the factor $1/\sqrt{p_1p_2p_3}$ in Eq.~\eqref{eq:C_BI} is quantized using a Tikhonov-type regularization (as for instance in \cite{Wilson-Ewing:2015xaq}) instead of a regularization based on the use of Thiemann identities.

\section{Conclusions}

In the present work we introduced, in the kinematical setting of loop quantum gravity, a master constraint operator which represents a set of gauge conditions fixing the densitized triad to be diagonal, and which therefore provides an alternative way of carrying out the quantum gauge fixing procedure underlying quantum-reduced loop gravity. We showed that the standard Hilbert space of quantum-reduced loop gravity can be recovered as a particular space of (approximate) solutions of the master constraint operator. We also argued that the standard reduced Hilbert space may not be sufficiently large to encompass all relevant configurations in which the chosen gauge is satisfied, and proposed an extension of the Hilbert space, which may be suitable to describe geometries in which the densitized triad is diagonal while the Ashtekar connection also has non-vanishing off-diagonal components.

As a first step towards analyzing the quantum dynamics of states in the quantum-reduced framework, we considered the dynamics of very simple reduced states, in which a single six-valent node represents the spatial geometry of a universe with the topology of a torus. Restricting our attention (for now) only to the Euclidean part of the Hamiltonian constraint, we derived the Hamiltonian which determines the dynamics of such states in the quantum-reduced model, and observed that the resulting operator exhibits a clear formal similarity with the Hamiltonian constraint of Bianchi I models in loop quantum cosmology.

An apparent drawback of the aforementioned Hamiltonian is that in the viewpoint of loop quantum cosmology, it corresponds to a constant value of the polymerization parameter $\mu$, i.e.~to the so-called $\mu_0$-scheme. At the moment it is not clear how, or if, the ``improved dynamics'' \cite{Ashtekar:2006wn} of the $\bar\mu$-scheme could be obtained from our construction, although we may note that at the level of effective dynamics, Alesci and Cianfrani have argued that the $\bar\mu$-scheme can be recovered from quantum-reduced loop gravity by considering a state represented by a density matrix, where the states in the ensemble correspond to graphs having different number of nodes \cite{Alesci:2016rmn}.

\section*{Acknowledgments}

This work was funded by National Science Centre, Poland, grant no. 2022\slash 44\slash C\slash ST2\slash 00023. For the purpose of open access, the author has applied a CC BY 4.0 public copyright license to any author accepted manuscript (AAM) version arising from this submission.

\printbibliography

\end{document}